\documentclass[draftcls,onecolumn]{IEEEtran}

\IEEEoverridecommandlockouts
\usepackage{booktabs}
\usepackage{float}
\usepackage{comment}
\usepackage{caption}
\usepackage{cite}
\usepackage{url}
\usepackage{amsmath,amssymb,amsfonts}
\usepackage{algorithmic}
\usepackage{graphicx}
\usepackage{textcomp}
\usepackage{xcolor}
\def\BibTeX{{\rm B\kern-.05em{\sc i\kern-.025em b}\kern-.08em
    T\kern-.1667em\lower.7ex\hbox{E}\kern-.125emX}}
\usepackage{hyperref}
\hypersetup{
    colorlinks=true,
    linkcolor=blue,      % Colors internal links (Figures, Tables, Equations)
    citecolor=blue,   % Colors citation links (\cite)
    urlcolor=blue        % Colors external URLs
}

\makeatletter
\def\bstctlcite{\@ifnextchar[{\@bstctlcite}{\@bstctlcite[\@nil]}}
\def\@bstctlcite[#1]#2{\@bsphack
  \@for\@citeb:=#2\do{%
    \edef\@citeb{\expandafter\@firstofone\@citeb}%
    \if@filesw\immediate\write\@auxout{\string\citation{\@citeb}}\fi}%
  \@esphack}
\makeatother
    
\begin{document}
\bstctlcite{IEEEexample:BSTcontrol}
%\onecolumn
\title{QmDFT for Polycyclic Aromatics: Balancing Embedding Ground-State Fidelity and Experimental Gap Estimation}

\author{
\IEEEauthorblockN{
N. Manglani$^{1,2}$\IEEEauthorrefmark{1},
Tejjan Arora$^{3}$,
Samrit Maity$^{3}$,
Ranjit Thapa$^{4}$,
Shashank Sharma$^{3}$,
Sanjay Wandhekar$^{3}$
}

\IEEEauthorblockA{
$^{1}$AICTE Industry Fellow, Centre for Development of Advanced Computing (C-DAC), Pune, India\\
$^{2}$Assistant Professor, Shah and Anchor Kutchhi Engineering College, Mumbai, India
}

\IEEEauthorblockA{
$^{3}$Centre for Development of Advanced Computing (C-DAC), Pune, India
}

\IEEEauthorblockA{
$^{4}$Department of Physics \& Centre for Computational and Integrative Sciences, SRM University - AP, Amaravati 522240, Andhra Pradesh, India
}

\IEEEauthorblockA{
\IEEEauthorrefmark{1}Corresponding author: \texttt{namrata.manglani@sakec.ac.in}
}
}

\maketitle

\begin{abstract}
 Quantum-Embedding density functional theory (QmDFT) embedding offers a highly scalable approach to improve treatment for large highly correlated $\pi$-conjugated systems. However, estimating advanced electronic structure properties in polycyclic aromatic hydrocarbons (PAHs) needs advanced exchange–correlation functionals that frequently trigger convergence instabilities during the embedding cycle.

In this work, we introduce an adaptive damping and direct inversion in the iterative subspace (DIIS)-accelerated protocol that stabilizes the embedding procedure, enabling robust integration of hybrid functionals like B3LYP and CAM-B3LYP. Using 10 selected PAHs (linear and fused) molecules as a benchmark. We demonstrate a clear functional-dependent ground-state energetics and frontier-orbital gap estimation. While LDA-based approaches yield near-quantitative agreement with FCI-in-DFT reference energies and is further supported by thermochemical isomerization benchmarks, while B3LYP provide significantly improved agreement with experimental $E_{0-0}$ transition values. This mapping allows us to bypass explicit excited-state calculations for $E_{0-0}$ values, thereby significantly reducing computational overhead.

Among the hybrid functionals screened, CAM-B3LYP offers a balanced overall performance. Our results establish a stable  QmDFT framework and provide useful guidance for functional selection for quantum embedding studies of PAHs and related low-dimensional $\pi$-conjugated materials.

\end{abstract}


\section{Introduction}

Accurate description of extended conjugated systems remains difficult because their electronic structure is strongly influenced by electron correlation and non-local exchange effects, which play a central role in low-dimensional carbon nanostructures \cite{Hachmann2007acenes, Kronik2012, Henderson2008}. Density Functional Theory (DFT) remains one of the most widely used approaches for materials modeling in such cases but its performance can deteriorate for extended $\pi$-conjugated hydrocarbons. High-accuracy methods such as full configuration interaction (FCI) can describe these effects, but their computational cost increases exponentially with system size, restricting their application to relatively small molecular systems. Quantum-Embedding density functional theory (QmDFT) embedding offers an efficient alternative that is computationally less demanding with polynomial scaling \cite{ Manby2012, Knizia2013,  Rossmannek2023, Manglani2026}.

The estimation ability of QmDFT depends on the choice of functional in DFT bath (Keeping quantum approach fixed). Historically across Jacob's ladder of density functionals, local density approximation (LDA) functionals have been known to yield satisfactory ground-state properties but severely underestimate electronic bandgaps due to self-interaction errors \cite{Perdew1981}, while B3LYP generally improves the estimation of excitation energies due to the inclusion of a fraction of exact exchange \cite{Becke1988, becke1993density}. While range-separated hybrids like CAM-B3LYP offer a balance by adjusting short- and long-range exact exchange \cite{yanai2004cam}. This behavior of functionals within an iterative embedding potential loop across different molecular topologies is relatively unexplored. Moreover, implementing higher correlation functionals in embedding self-consistent cycles raises numerical challenges which have not been explored earlier for PAH like strongly correlated systems.

To explore the unmapped territory the present work introduces an adaptive damping and DIIS-based convergence protocol for stabilizing range-separated hybrid loops against numerical convergence failures within the QmDFT framework. We study ten structurally diverse PAHs under this framework and establish inverse error relationship. Range-separated LDA (LDA-RS) that performs best in embedding gives lowest error compared to FCI-in-DFT baseline while B3LYP functional minimizes $E_{0-0}$ gap errors. Finally we demonstrate that anchoring approach—calibrating the CAM-B3LYP exchange parameters ($\alpha+\beta$) to a single reference system (anthracene)—effectively bridges this gap, compressing both ground-state and electronic transition errors simultaneously across the low-dimensional series.

\section{Theory }\label{sec:theory}

\subsection{Projection-Based DFT Quantum Embedding}\label{subsec:projection_embedding}

Projection-based density functional theory (DFT) embedding~\cite{Manby2012, Manglani2026}
couples a high-level wavefunction method (here, variational quantum eigensolver, VQE) within an
active space to a low-level DFT treatment of the environment. The total energy is decomposed as

\begin{equation}
E_\text{total} = E_\text{env}[\rho_\text{inact}] + E_\text{act}[\rho_\text{act}] + E_\text{emb}[\rho_\text{act}, \rho_\text{inact}]
\label{eq:embedding_energy}
\end{equation}

where \(\rho_\text{inact} = \rho_\text{DFT}^\text{full} - \rho_\text{act}\) is the inactive (environment) density
constructed from the DFT calculation on the full system, and \(\rho_\text{act}\) is the active‑region
density obtained from the VQE solver. The term \(E_\text{emb}\) encapsulates the nonadditive kinetic and
Coulomb contributions that couple the active region to the environment.

In the present implementation, self-consistency is achieved by iteratively updating the active density
\(\rho_\text{act}^{(n)}\) and recomputing the DFT environment from the total density
\(\rho_\text{total}^{(n)} = \rho_\text{inact} + \rho_\text{act}^{(n)}\). The embedding loop may be summarized as

\begin{equation}
\rho_\text{act}^{(n+1)} = \mathcal{P}_\text{act} \left[
\psi_\text{VQE} \left( H_\text{emb}[\rho_\text{inact}, \rho_\text{act}^{(n)}] \right)
\right]
\label{eq:projection_loop}
\end{equation}

where \(\psi_\text{VQE}\) is the VQE solution within the active space, \(H_\text{emb}\) is the embedding
Hamiltonian constructed from the environment Fock operator \(F_\text{env}\) evaluated at the updated
total density, and \(\mathcal{P}_\text{act}\) is a projection onto the active subspace that prevents
double occupancy between the DFT environment and the VQE solution. In practice, the projection is
enforced by limiting the active density to the chosen active orbitals and by damping the density updates
between iterations, which stabilizes the fixed‑point convergence for difficult systems~\cite{Manby2012, Rossmannek2021, Rossmannek2023 } . 

\begin{figure}[htbp]
\centering
\includegraphics[width=0.9\textwidth]{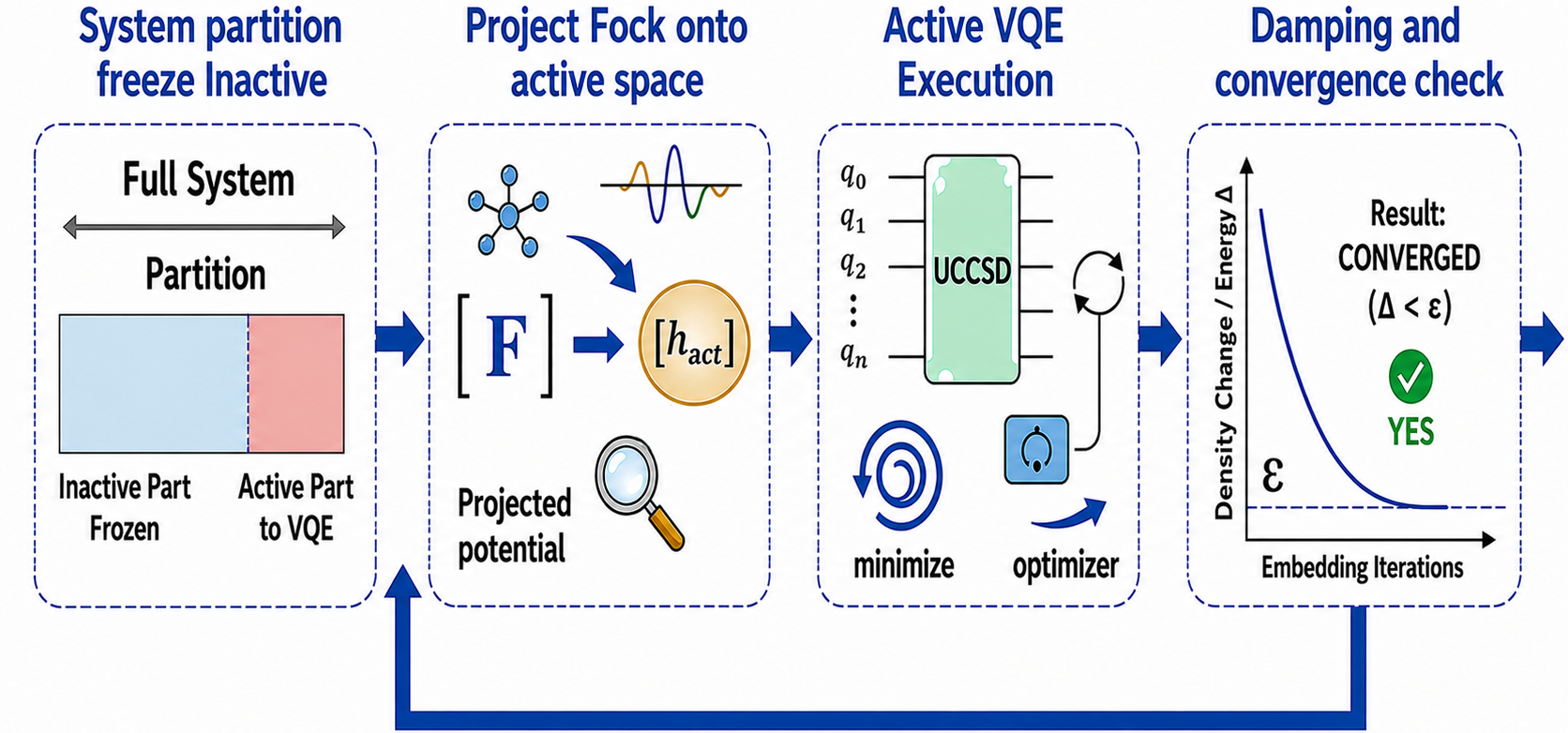}
\caption{QmDFT Schematic diagram}
\label{fig:QmDFT_schematic}
\end{figure}

As shown in Fig. \ref{fig:QmDFT_schematic} 
our implementation follows the general spirit of projection‑based WF‑in‑DFT and VQE‑in‑DFT embedding schemes introduced in the literature~\cite{Manby2012,  Rossmannek2021}, while relying on a practical iterative mixing protocol rather than an explicit level‑shift projector on the occupied KS manifold.

\subsection{Advanced Functionals and Convergence Control in QmDFT}\label{subsec:higher_functionals_and_convergence}

The projection-based embedding framework utilized in this work provides a natural generalization of the Quantum-Embedding density functional theory (QmDFT) philosophy~\cite{Manglani2026}: a small active region is treated at a high-level (here VQE), while the environment is represented by a computationally efficient DFT-based bath. The low-level functional is initially chosen as LDA-RS, which yields stable embedding loops due to its short-range dominance. However, for the accurate estimation of frontier-orbital gaps in extended $\pi$-conjugated systems such as polycyclic aromatic hydrocarbons (PAHs), longer-range exchange effects must be captured~\cite{yanai2004cam}. This necessitates the use of more advanced functionals, including range-separated hybrids (RSH) such as CAM-B3LYP.

The Standard Trotterized UCCSD ansatz used in the VQE part has a problem that it tends to break the $\hat{S}^2$ symmetry mostly during the use of advanced functionals. This leads to triplet contamination in the target singlet ground-state. To reduce this effect, a spin-penalty approach was employed following Kuroiwa and Nakagawa~\cite{Kuroiwa2021}. The modified Hamiltonian is written as: $\hat{H}_{vqe} = \hat{H} + \beta_{spin} \hat{S}^2$, where the penalty coefficient is defined as $\beta_{spin} = \Delta E_{ST} / C_{min}^2$ ~\cite{Kuroiwa2021} with $\Delta E_{ST}$ denoting the  singlet--triplet energy gap and $C_{\min}^{2}=0.5625$ for the spin-squared operator. This theoretical threshold ensures that we reject higher-multiplicity states.

Hybrid and range-separated hybrid functionals can introduce oscillatory behavior during density mixing within the embedding cycle. To stabilize the optimization trajectory, a two-stage "stabilize-then-accelerate" procedure was employed. During the initial stage, the active-space density matrix was updated according to
\begin{equation}
\rho_{\text{act}}^{\text{damped}} = (1-\alpha_k)\rho_{\text{act}}^{(k-1)} + \alpha_k \rho_{\text{act}}^{(k)}
\label{eq:linear_damping}
\end{equation}
where $\rho_{\text{act}}^{(k)}$ denotes the raw active density matrix computed at iteration $k$ and $\alpha_k$ acts as a dynamically decaying mixing coefficient. To suppress high-frequency fluctuations early while ensuring the system can step deeply into the convergence manifold, the parameter is scaled inversely with the progress of the sequence:
\begin{equation}
\alpha_k \propto \frac{\alpha_0}{\sqrt{k}}
\label{eq:damping_decay}
\end{equation}
where $\alpha_0$ is the fixed initial mixing parameter. This damping scheme was applied during the early stages of the embedding cycle, before accelerated subspace extrapolation methods can safely take over.

\subsection{Coulomb Attenuation Framework for Asymptotic Range-Separation}

Long-range exchange interactions were described using the range-separated hybrid functional CAM-B3LYP. In this functional, the electron--electron Coulomb operator is partitioned according to the Coulomb attenuation scheme~\cite{yanai2004cam}:
\begin{equation}
\frac{1}{r_{12}} = \frac{\alpha+\beta\,\text{erf}(\mu r_{12})}{r_{12}} + \frac{1-\left(\alpha+\beta\,\text{erf}(\mu r_{12})\right)}{r_{12}},
\label{eq:coulomb_attenuation}
\end{equation}
where $\mu$ is the range-separation parameter and $\alpha$ and $\beta$ determine the relative contributions of short- and long-range exchange. This approach significantly improves charge‑transfer and polarization descriptions in delocalized networks compared to conventional global hybrids or pure generalized gradient approximations~\cite{Kronik2012, Chen2019}.

\section{Computational Methods}
\label{sec:methods}

\subsection{Quantum DFT Embedding Framework}

All calculations employed projection-based DFT embedding within the QmDFT formalism, partitioning PAHs into $\pi$-active spaces embedded in DFT-generated baths~\cite{Manby2012, Manglani2026}. Computations utilized \texttt{PySCF} 2.5+~\cite{pyscf2023} interfaced with \texttt{Qiskit Nature} 0.7+~\cite{QiskitNature2024} on the C-DAC PARAM Shakti HPC Facility. 

Embedding baths used functionals ranging from LDA-RS to range-separated hybrids, while the active regions utilized 6-31G* basis sets. Convergence for higher-rung functionals was achieved via Hartree--Fock VQE initialization with second-order M{\o}ller--Plesset perturbation theory (MP2)~\cite{Romero2019}. The embedding self-consistency convergence tolerance ($\epsilon_\text{tol}$) was strictly set to $10^{-6}$ Ha \textit{and} the Frobenius norm of the active density matrix difference ($\Delta \rho_{Frob}$) was set to $10^{-4}$.

\subsection{Implementation of Adaptive Damping and DIIS}

Our custom \texttt{DFTEmbeddingSolver} implements a dynamic, two-stage convergence protocol to stabilize the iterative updating of the active density matrix, considering the self-consistency cycle fully converged when the energy variance drops below $10^{-6}$~Ha and the Frobenius norm of the active density matrix difference ($\Delta \rho_{\text{Frob}}$) settles below $10^{-4}$. To balance early-stage robustness with late-stage acceleration, this operational profile is divided into two sequential phases:\\

\textbf{Phase I: Adaptive Linear Damping:}

\begin{figure}[H]
\centering
\includegraphics[width=0.9\textwidth]{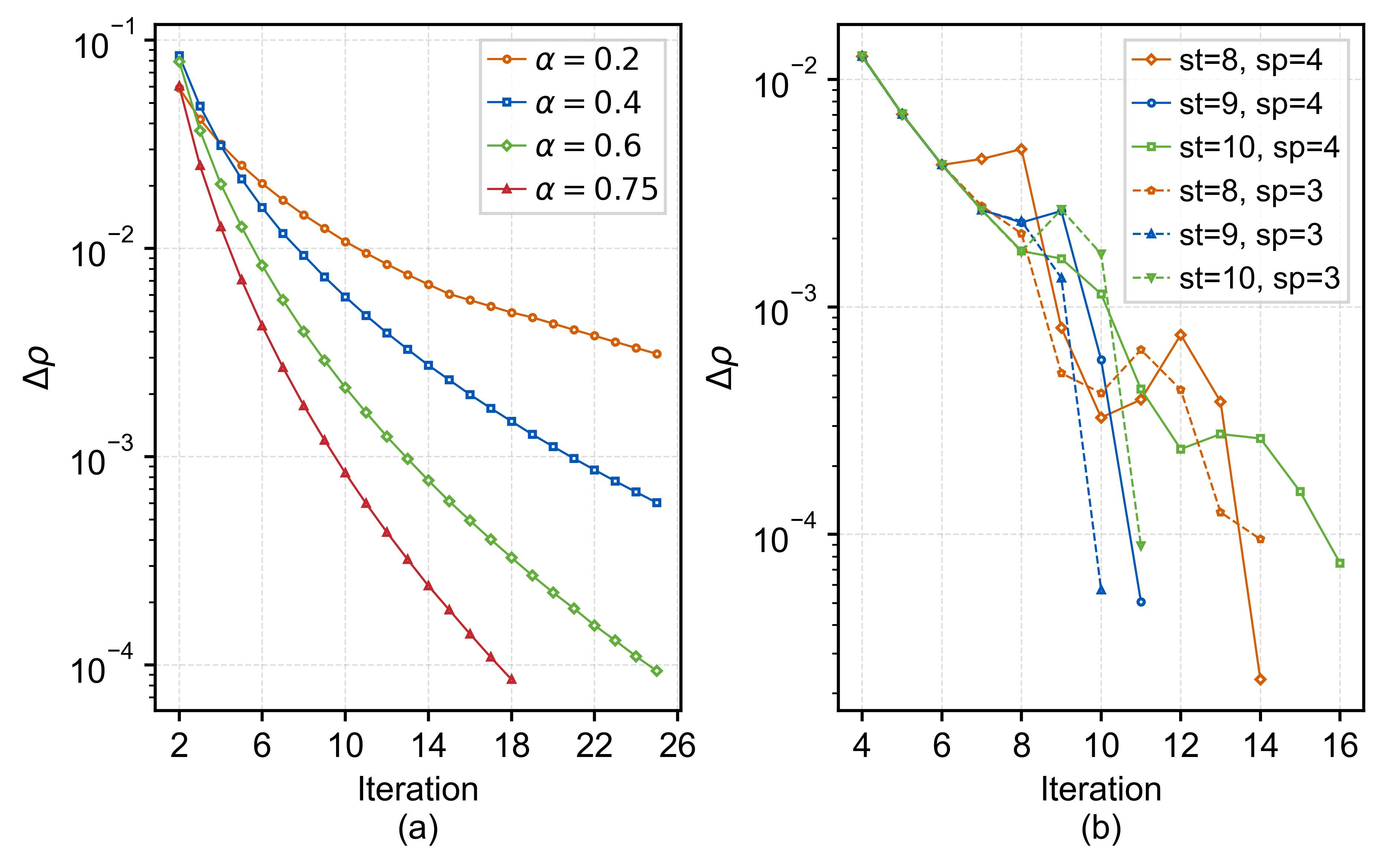}
\caption{Convergence analysis for anthracene using the CAM-B3LYP functional. (a) Density convergence for different damping parameters $\alpha$. (b) Convergence obtained with different DIIS activation steps ($st$) and history sizes ($sp$). The selected parameters correspond to $\alpha=0.75$, $st=9$, and $sp=3$.}
\label{fig:damping}
\end{figure}

During the initial SCF iterations, damping was applied to the active-space density matrix to reduce oscillatory behavior~\cite{Epifanovsky2021}. The density update was performed according to
\begin{equation}
\rho_\text{act}^\text{damped} = (1-\alpha_k) \rho_\text{act}^{(k-1)} + \alpha_k \rho_\text{act}^{(k)}
\label{eq:linear_damping}
\end{equation}
where the damping parameter was reduced as $\alpha_k \propto 1/\sqrt{k}$. The value of $\alpha$ was selected by comparing the resulting density convergence. When multiple values produced similar $\Delta\rho$, the parameter requiring fewer iterations was chosen. Figure~\ref{fig:damping}(a) presents results for anthracene with the CAM-B3LYP functional, where $\alpha=0.75$ produced the fastest convergence to the $\Delta\rho < 10^{-4}$ criterion. Since comparable behavior was observed for the remaining functionals, $\alpha=0.75$ was used throughout this work.

\textbf{Phase II: DIIS Acceleration:}

DIIS extrapolation~\cite{Pulay1980} is activated after the initial damping iterations. We benchmarked different combinations of the DIIS activation iteration and history subspace size, as shown in Figure~\ref{fig:damping}(b). The benchmarks reveal that activating the DIIS routine at the 9th iteration with a subspace history of the three most recent density matrices ($st=9, sp=3$) yields the cleanest, most efficient convergence path, avoiding the minor stability oscillations seen with larger subspaces.

\subsection{Active Space and Quantum Solver Protocol}\label{subsec:quantum_solver}

Active spaces between $(2e, 6o)$ and $(8e, 6o)$ were considered to describe the primary $\pi \rightarrow \pi^*$ excitations in the acene series. Comparison across these active spaces enabled an assessment of the trade-off between computational cost and correlation treatment. Based on the active-space analysis, an $(6e,6o)$ active space was selected for the remaining calculations.

While a six-orbital active space may not capture the full extent of correlation effects in larger acenes such as pentacene, it provides a uniform framework for comparison across all systems and functionals considered in this work.

The active-space Hamiltonian, $H_{\mathrm{emb}}[\rho^{(n)}]$, was solved using the variational quantum eigensolver (VQE) with a Unitary Coupled Cluster Singles and Doubles (UCCSD) ansatz ($\mathrm{reps}=1$). The fermionic Hamiltonian was converted to a qubit Hamiltonian using parity mapping with subsequent $C_{2v}$ symmetry tapering, resulting in a two-qubit reduction~\cite{Bravyi2017}.

A single-step Trotterized UCCSD ansatz was generated from the Hartree--Fock reference state. Variational optimization was performed with L-BFGS-B using MP2-derived initial amplitudes~\cite{Romero2019}. Convergence of the VQE optimization was defined by a relative energy threshold of $\Delta E < 10^{-6}$ Ha and Frobenius norm of the density threshold $\Delta \rho_{Frob} < 10^{-4}$, with a maximum of 100 iterations~\cite{Byrd1995}.

All simulations were performed on a noiseless statevector backend to isolate the intrinsic performance of the embedding and functional framework from hardware-induced noise and sampling errors.

\subsection{Molecular Geometries and Evaluation Protocol}

\begin{figure}[htbp!]
\centering
\includegraphics[width=0.9\textwidth]{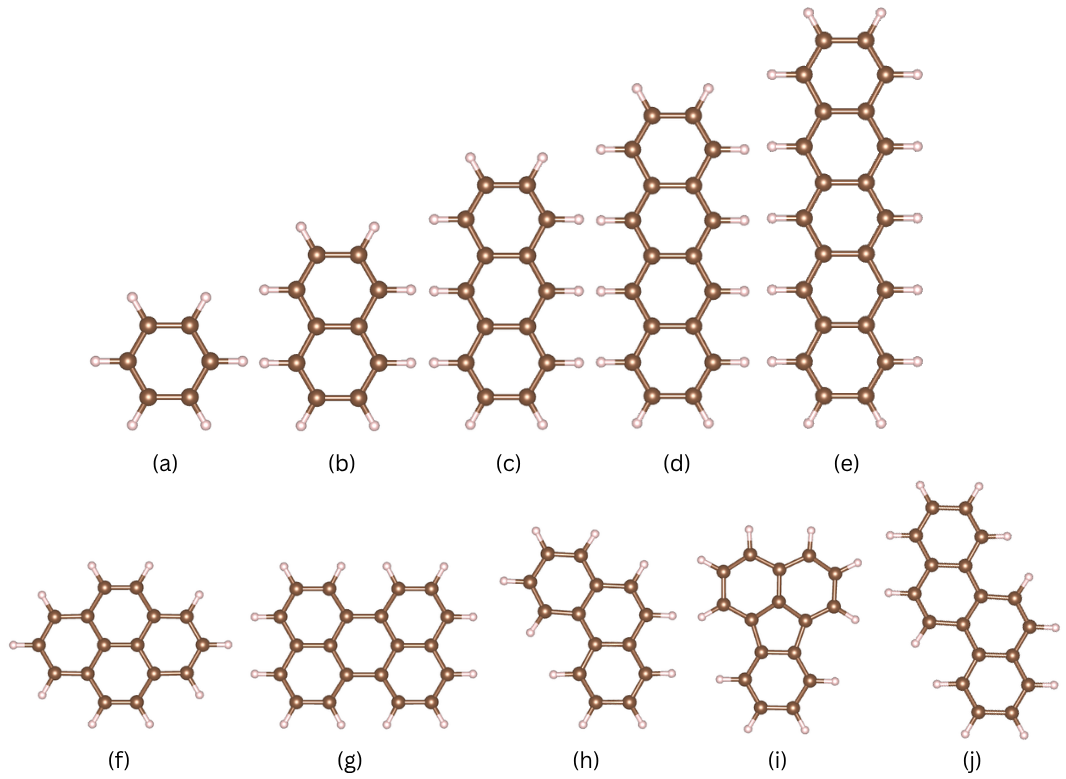}
\caption{Molecular ball-and-stick models of the PAHs utilized for QmDFT benchmarking : (a) benzene ($C_6H_6$), (b) naphthalene ($C_{10}H_8$), (c) anthracene ($C_{14}H_{10}$), (d) tetracene ($C_{18}H_{12}$), (e) pentacene ($C_{22}H_{14}$), (f) pyrene ($C_{16}H_{10}$), (g) perylene ($C_{20}H_{12}$), (h) phenanthrene ($C_{14}H_{10}$), (i) fluoranthene ($C_{16}H_{10}$), and (j) chrysene ($C_{18}H_{12}$). Geometries optimized at the B3LYP/6-31G* level; structural visualizations rendered via VESTA.}
\label{fig:vesta-pah}
\end{figure}

The study focuses on the PAHs comprising benzene (C$_6$H$_6$), naphthalene (C$_{10}$H$_8$), anthracene (C$_{14}$H$_{10}$), tetracene (C$_{18}$H$_{12}$), pentacene (C$_{22}$H$_{14}$), pyrene ($C_{16}H_{10}$), perylene ($C_{20}H_{12}$), phenanthrene ($C_{14}H_{10}$), fluoranthene ($C_{16}H_{10}$), and chrysene ($C_{18}H_{12}$). Molecular structures obtained from vesta are shown in Fig. \ref{fig:vesta-pah}. Initial molecular structures were obtained from the NIST Computational Chemistry Comparison and Benchmark Database (CCCBDB) and subsequently refined via geometry optimization at the B3LYP/6-31G* level using the \texttt{PySCF} electronic structure package interfaced with the geomeTRIC optimization engine.

We used these optimized geometries as the consistent starting point for all subsequent embedding calculations. To evaluate the performance of the QmDFT (VQE-in-DFT) framework, we applied a series of energy- and property-based tests to this structural dataset. Total electronic energies obtained from the VQE-in-DFT calculations were compared with corresponding FCI-in-DFT reference values under the same embedding conditions. The absolute energy deviation which represents the Variational optimization error here, $ |\Delta E| = |E_{\text{QmDFT}} - E_{\text{FCI-in-DFT}}| $, was used to evaluate the Mean Absolute Error (MAE) and Root-Mean-Square Error (RMSE) for 10 selected PAHs. 

As an additional validation of the energy being used for property estimations, relative isomerization energies were calculated for the anthracene/phenanthrene and tetracene/chrysene pairs in QmDFT using different functionals. These were then compared against experimental reference values were calculated by taking the difference between the respective experimental gas-phase enthalpies of formation ($\Delta_f H^\circ_{\mathrm{gas}}$) obtained from Roux et al.\cite{Roux2008} to assess each exchange--correlation functional's relative energetic ordering.

To evaluate electronic spectral properties, static ground-state Kohn--Sham (KS) frontier-orbital gaps ($E_{\text{HL}}$) were evaluated as a direct proxy for the experimental $E_{0-0}$ optical transition energies. This property matching was conducted across two distinct regimes to quantify the explicit benefit of the quantum embedding loop:

\begin{itemize}
\item \textbf{Embedded QmDFT gap Error Profile:} The gap error was evaluated using the embedded frontier eigenvalues extracted directly from the self-consistent active density space:
$$ \Delta E_{\text{gap}}^{\text{QmDFT}} = |E_{\text{HL}}^{\text{QmDFT}} - E_{0-0}| $$
\item \textbf{ DFT gap Error Profile:} The gap error was evaluated using the pure, Plain DFT eigenvalues across identical functional environments:
$$ \Delta E_{\text{gap}}^{\text{DFT}} = |E_{\text{HL}}^{\text{DFT}} - E_{0-0}| $$
\end{itemize}

Statistical error distributions ($\text{MAE}_{\text{gap}}$ and $\text{RMSE}_{\text{gap}}$) were calculated both regimes across all five exchange--correlation functionals. Comparing $\text{MAE}_E$ against $\text{MAE}_{\text{gap}}$ provides the mathematical basis for mapping the inverse error relationship between ground-state energy precision and orbital energy errors. Furthermore, tracking the differential performance between $\Delta E_{\text{gap}}^{\text{QmDFT}}$ and $\Delta E_{\text{gap}}^{\text{DFT}}$ enables quantification of the embedding efficiency for each functional.

\section{Results}
\label{sec:results}

\subsection{Global Functional Anchoring and Parameter Calibration}
To model extended $\pi$-conjugated low-dimensional networks reliably, the range-separated hybrid functional parameters must be anchored to a representative physical system. In this work, we maintain the standard default short-range parameters of CAM-B3LYP ($\mu = 0.33$ and $\alpha = 0.19$), but treat the long-range parameter $\beta$ as an explicit control variable to calibrate the behavior of the single-particle ground-state Kohn–Sham gaps against experimental trends.

Using anthracene (experimental $E_{0-0}$ gap target of $3.402\text{ eV}$~\cite{Benkyi2019}) as our representative reference evaluated within the uniform $(6e, 6o)$ active space, our parameter scan demonstrates that the embedded QmDFT HOMO--LUMO gap scales linearly and monotonically with the long-range exchange fraction $\beta$. As shown in our calibration profile, increasing $\beta$ expands the embedded orbital gap from $3.244\text{ eV}$ at $\beta = 0.05$ up to a maximum of $4.863\text{ eV}$ at $\beta = 0.41$. A closest proximity to the experimental target is provided with $\beta = 0.09$ (yielding a combined global exchange threshold of $\alpha+\beta = 0.28$), which provides gap calculation of $3.384\text{ eV}$. This calibrated configuration, denoted as CAM-B3LYP*, is utilized as the uniform baseline for all subsequent cross-topology evaluations.

\subsection{Active-space sensitivity analysis}
 To evaluate sensitivity of the accuracy on active space selection, energy deviations between QmDFT and FCI-in-DFT were calculated for the LDA-RS functional for four active spaces $(2e,6o)$, $(4e,6o)$, $(6e,6o)$ and $(8e,6o)$ for the five molecules in acene series. The absolute deviations for all twenty possibilities are summarized in Table~\ref{tab:correlation_errors}.

\begin{figure}[htbp!]
\centering
\includegraphics[width=0.9\textwidth]{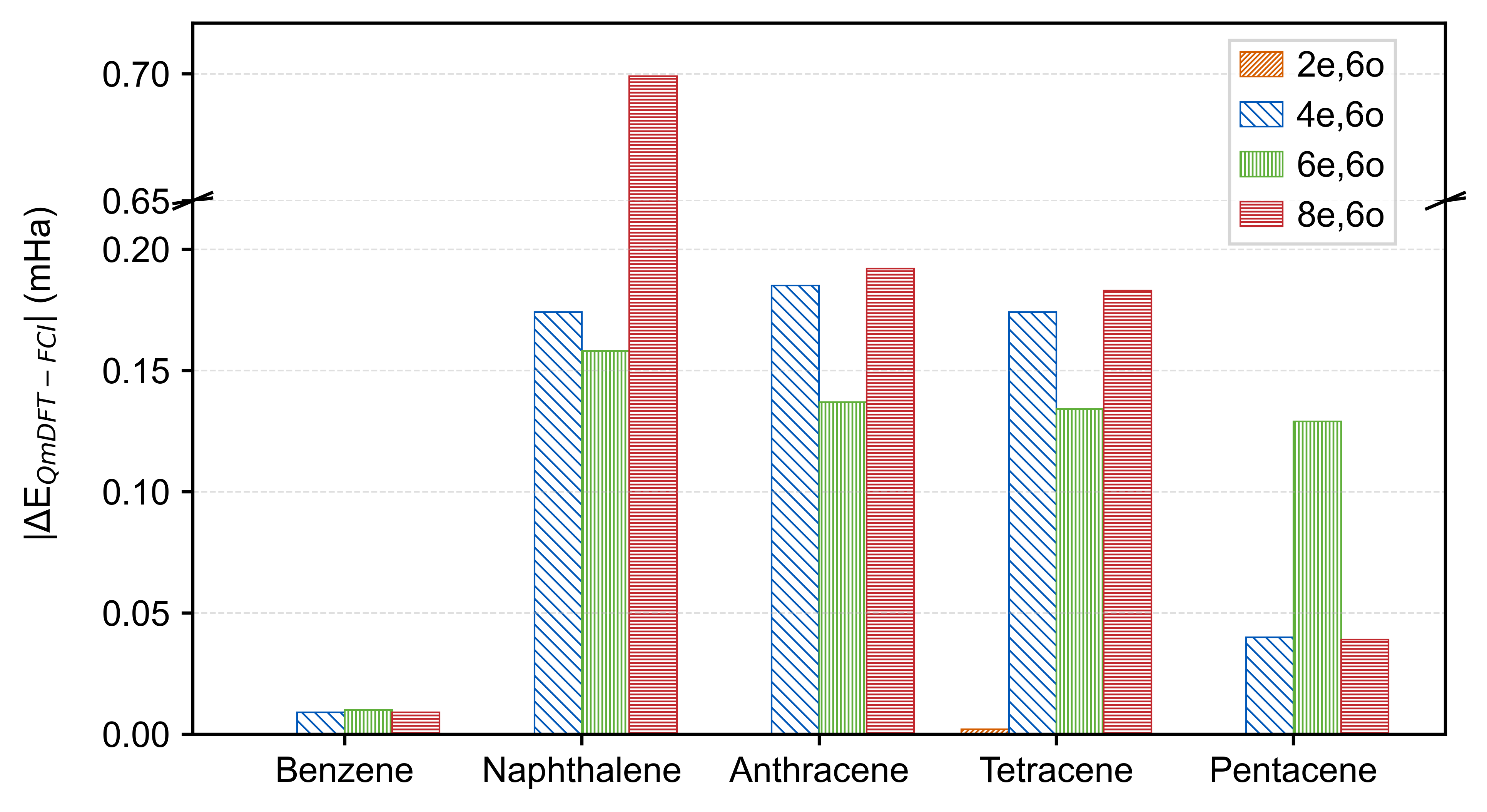}
\caption{Absolute energy deviations ($|\Delta E|$, mHa) between QmDFT and FCI-in-DFT calculations for 10 selected PAHs using the LDA-RS functional across different active spaces. The overall MAE and RMSE are 0.118 mHa and 0.192 mHa, respectively.}
\label{fig:deltaE}
\end{figure}

\begin{table}[htbp]
\centering
\caption{Absolute correlation energy deviations $|\Delta E|$ (mHa) between QmDFT and FCI-in-DFT calculations for 10 selected PAHs using the LDA-RS functional.}
\label{tab:correlation_errors}
\footnotesize
\begin{tabular}{lcccc}
\hline
Molecule & $(2e,6o)$ & $(4e,6o)$ & $(6e,6o)$ & $(8e,6o)$ \\
\hline
Benzene     & 0.000 & 0.009 & 0.010 & 0.009 \\
Naphthalene & 0.000 & 0.174 & 0.158 & 0.699 \\
Anthracene  & 0.000 & 0.185 & 0.137 & 0.192 \\
Tetracene   & 0.002 & 0.174 & 0.134 & 0.183 \\
Pentacene   & 0.000 & 0.040 & 0.129 & 0.039 \\
\hline
\end{tabular}
\end{table}

As shown in Table~\ref{tab:correlation_errors} and Fig.~\ref{fig:deltaE} the absolute total energy deviations of QmDFT from FCI-in-DFT remain below 0.20 mHa for most active spaces considered. The smallest active space $(2e,6o)$ produces  smallest energy deviations. This primarily reflects the limited correlation complexity of this compact active-space Hamiltonian which is not able to adequately capture the dominant $\pi$-electron correlation effects across the acene series.

As the active space increases it is observed that it is sensitive to the electronic structure of the acene system. An anomalous deviation observed for Naphthalene $(8e,6o)$ likely originates from the increased difficulty of approximating the correlated active-space Hamiltonian within the QmDFT framework.

While, Pentacene exhibits larger deviations at the $(6e,6o)$ relative to other active spaces. This behavior reflects sharply increasing multireference charater from pentacene onwards in acene systems.

Among the active spaces considered, for Naphthalene, Anthracene, and Tetracene, the $(6e,6o)$ active space provides the best balance between correlation completeness, numerical stability, and agreement with FCI-in-DFT reference calculations. This motivates us to select this active space for subsequent cross-functional comparisons.

\subsection{HOMO--LUMO Gaps Across Functionals}

\begin{figure}[htbp]
\centering
\includegraphics[width=0.9\textwidth]{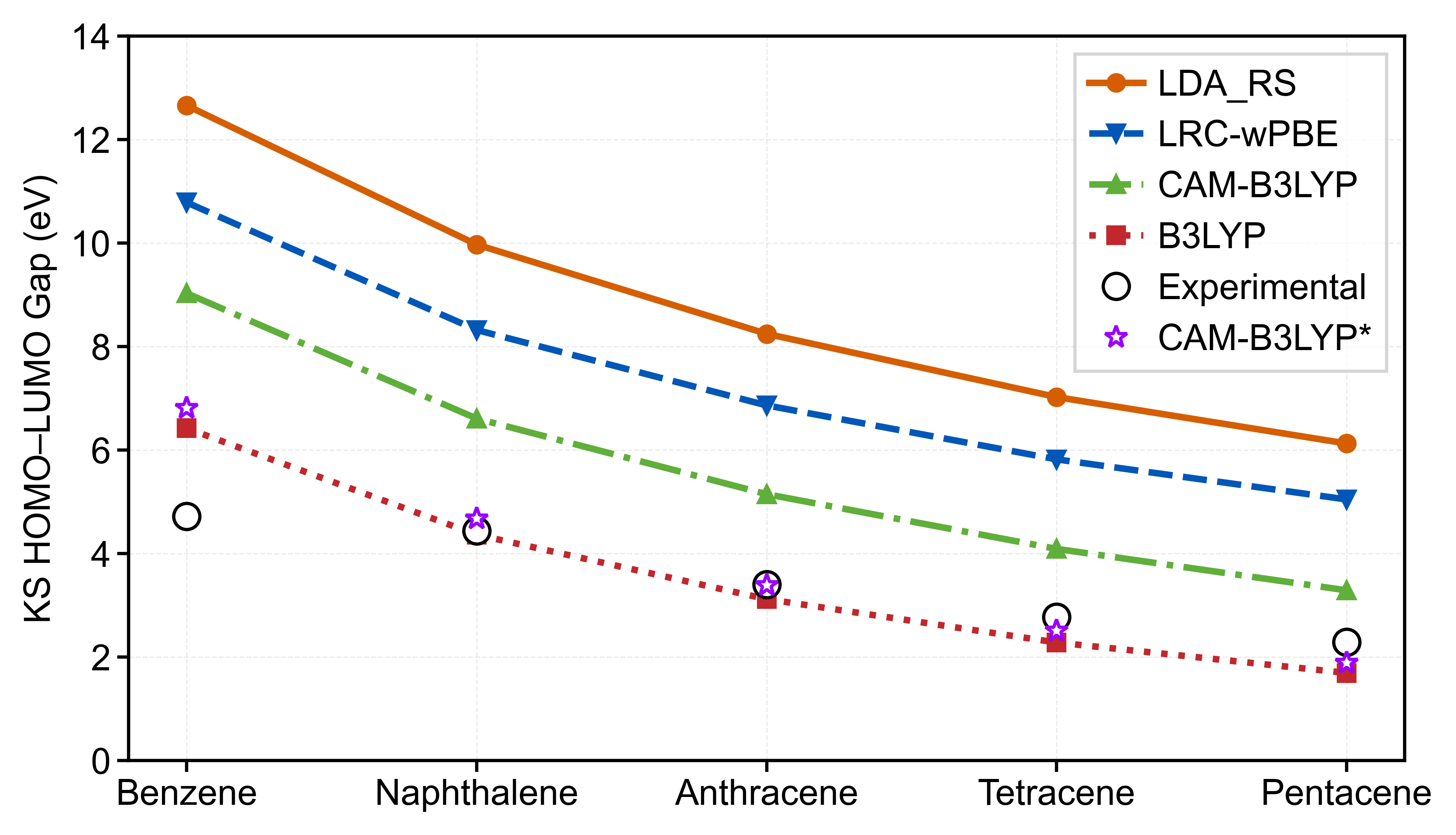}
\caption{HOMO--LUMO gaps of linear acenes computed using a fixed $(6e,6o)$ active space with LDA-RS, LRC-$\omega$PBE, CAM-B3LYP, B3LYP, tuned CAM-B3LYP and experimental reference values. (CAM-B3LYP* denotes tuned functional calibrated against anthracene.)}
\label{fig:gaps}
\end{figure}

HOMO--LUMO gaps were evaluated for the entire set of selected PAHs and plotted only for linear acenes from benzene through pentacene using the $(6e,6o)$ active space. Calculations were performed with the LDA-RS, LRC-$\omega$PBE, CAM-B3LYP, B3LYP and tuned CAM-B3LYP (CAM-B3LYP*) functionals. Numerical results for entire set of PAHs considered are presented in Table~\ref{tab:homo_lumo_gaps} along with experimental $E_{0-0}$ transition energy, and only the corresponding acene-series trends are displayed in Figure~\ref{fig:gaps}.

\begin{table}[htbp]
\centering
\caption{HOMO--LUMO gaps (eV) of linear  acenes computed using a fixed $(6e,6o)$ active space with different exchange--correlation functionals. Experimental references are included.}
\label{tab:homo_lumo_gaps}
\footnotesize
\begin{tabular}{lcccccc}
\hline
Molecule & LDA-RS & LRC-$\omega$PBE & CAM-B3LYP & B3LYP & CAM-B3LYP* & Experimental \\
\hline
Benzene      & 12.660 & 10.786 & 9.030 & 6.422 & 6.811 & 4.720 ~\cite{Callomon1966} \\
Naphthalene  & 9.972 & 8.321 & 6.610 & 4.366 & 4.674 & 4.443 ~\cite{Benkyi2019} \\
Anthracene   & 8.247  & 6.864 & 5.148 & 3.120 & 3.385 & 3.402 ~\cite{Benkyi2019} \\
Tetracene    & 7.025  & 5.821 & 4.091 & 2.282 & 2.507 & 2.775 ~\cite{Nina2013} \\
Pentacene    & 6.127  & 5.046 & 3.290 & 1.694 & 1.884 & 2.286 ~\cite{Benkyi2019} \\
Pyrene       & 8.525  & 7.197 & 5.616 & 3.493 & 3.788 & 3.842 ~\cite{Benkyi2019} \\
Perylene     & 7.588  & 6.526 & 4.923 & 2.734 & 3.034 & 2.983 ~\cite{Huisken2011} \\
Phenanthrene & 9.856  & 8.339 & 6.668 & 4.359 & 4.687 & 4.385 ~\cite{Huisken2011} \\
Fluoranthene & 8.969  & 7.689 & 6.070 & 3.679 & 4.035 & 3.127 ~\cite{Huisken2011} \\
Chrysene     & 9.197  & 7.828 & 6.186 & 3.890 & 4.212 & 3.496 ~\cite{Borisevich2008} \\
\hline
\end{tabular}
\end{table}

Across all functionals for QmDFT framework, the HOMO--LUMO gap decreases systematically from benzene to pentacene following an expected DFT trend. LDA-RS yields the largest gap values for all molecules, ranging from 12.660 eV (benzene) to 6.127 eV (pentacene). B3LYP produces lower gap values across the series, with values decreasing from 6.422 eV to 1.694 eV closely matching the experimental $E_{0-0}$ energy, except for Benzene. CAM-B3LYP followed by LRC-$\omega$PBE functional yields intermediate gap values between LDA-RS and B3LYP as shown in Figure~\ref{fig:gaps}.

For the tuned CAM-B3LYP (CAM-B3LYP*) functional which is tuned for anthracene and B3LYP functional, we find computed KS HOMO--LUMO gaps closely follow experimentally measured $E_{0-0}$ gaps.

\subsection{Error Profiles and the Inverse Functional Accuracy Relationship}

In order to assess the performance of functionals in QmDFT framework within the actice space $(6e,6o)$, a statistical error analysis was conducted. Table~\ref{tab:functional_errors} summarizes the Isomerization Energy Errors, the Mean Absolute Error (MAE) and Root-Mean-Square Error (RMSE) for total ground-state energies, together with the corresponding errors in both embedded (QmDFT) and DFT frontier-orbital gap estimates evaluated against experimental $E_{0-0}$ energy.

\begin{table}[htbp]
\centering
\caption{Comprehensive performance benchmarks across exchange--correlation functionals at the $(6e,6o)$ active space: Total ground-state energy deviations (mHa) relative to FCI-in-DFT, alongside QmDFT and DFT gap errors (eV) evaluated against experimental references. \textsuperscript{*}Denotes the globally anchored configuration calibrated against anthracene.}
\label{tab:functional_errors}
\small % Slightly larger than footnote for better Q1 readability
\begin{tabular}{lccccc}
\toprule
Metric & LDA-RS & LRC-$\omega$PBE & CAM-B3LYP & B3LYP & CAM-B3LYP\textsuperscript{*} \\
\midrule
\multicolumn{6}{c}{\textit{Isomerization Energy Errors ($|\Delta E_{\mathrm{iso}}-\Delta E_{\mathrm{exp}}|$, mHa)}} \\
A--P$^{a}$ & 3.466 & 5.892 & 9.589 & 14.771 & 14.515 \\
T--C$^{b}$ & 12.541 & 17.472 & 25.211 & 36.876 & 36.162 \\
\midrule
\multicolumn{6}{c}{\textit{Energy variational error($|E_{\text{QmDFT}} - E_{\text{FCI-in-DFT}}|$, mHa)}} \\
MAE  & 0.092 & 0.162 & 0.340 & 1.259 & 1.003 \\
RMSE & 0.102 & 0.181 & 0.391 & 1.329 & 1.074 \\
\midrule
\multicolumn{6}{c}{\textit{QmDFT Gap Errors ($|E_{\text{HL}}^{\text{QmDFT}} - E_{0-0}|$, eV)}} \\
MAE  & 5.271 & 3.896 & 2.217 & 0.471 & 0.504 \\
RMSE & 5.382 & 3.997 & 2.389 & 0.650 & 0.781 \\
\midrule
\multicolumn{6}{c}{\textit{ DFT Gap Errors ($|E_{\text{HL}}^{\text{DFT}} - E_{0-0}|$, eV)}} \\
MAE  & 5.520 & 4.347 & 2.817 & 0.479 & 0.871 \\
RMSE & 5.625 & 4.433 & 2.927 & 0.772 & 1.085 \\
\bottomrule
\end{tabular}
\vspace{1mm}

\raggedright
\footnotesize
$^{a}$ Anthracene--Phenanthrene; CCSD(T) deviation = 1.304 mHa.\\
$^{b}$ Tetracene--Chrysene; CCSD(T) deviation = 10.212 mHa.
\end{table}

\begin{figure}[htbp]
\centering
\includegraphics[width=0.9\textwidth]{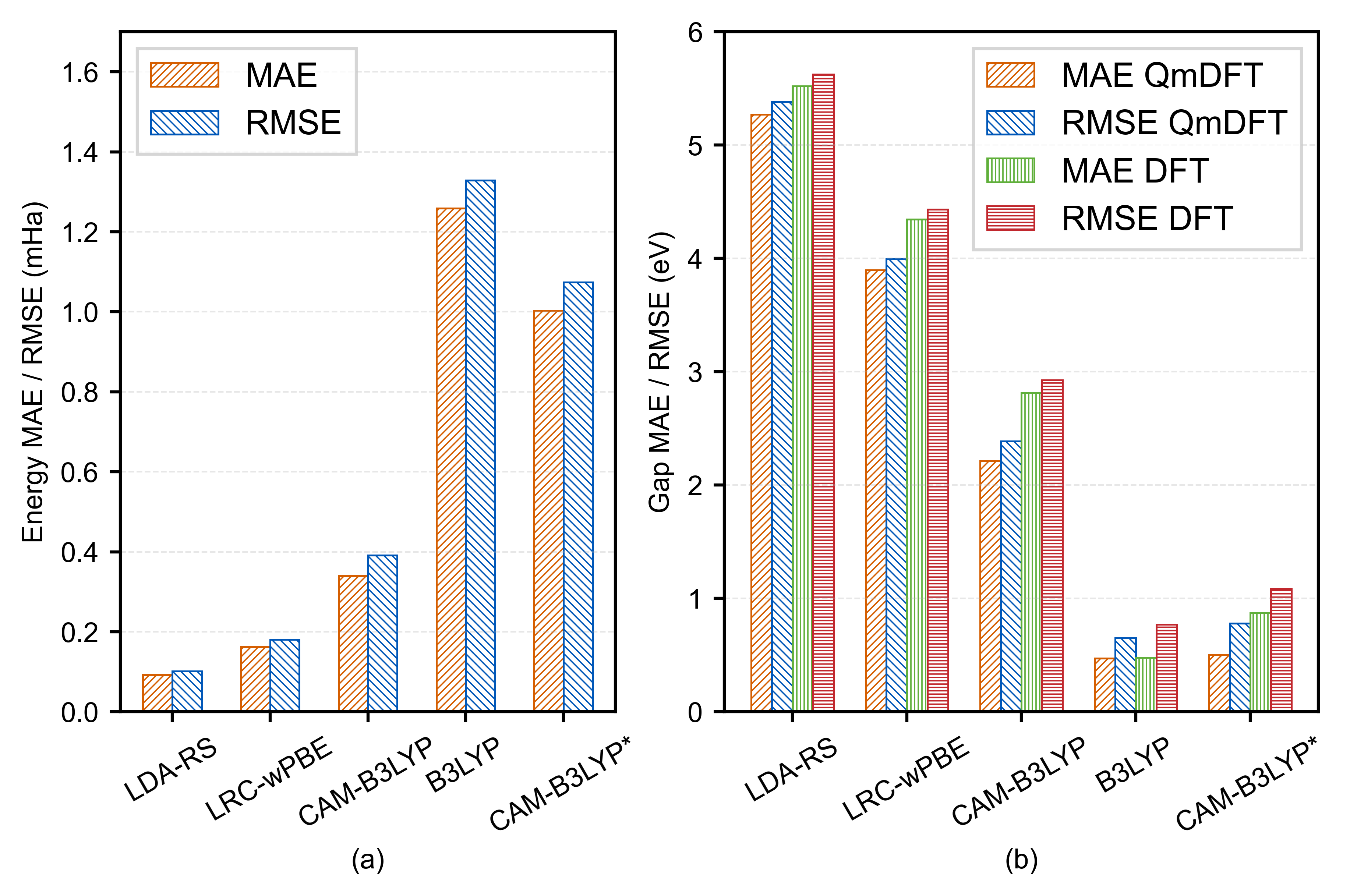}
\caption{Error analysis for the ten selected PAHs. (a) Energy deviations (mHa) from active-space$(6e,6o)$ vs corresponding FCI baselines. (b) Orbital gap errors (eV) evaluated against experiment.}
\label{fig:functional_perf}
\end{figure}

The error analysis demonstrates functional dependence of energetic and spectral observables as shown in Figure~\ref{fig:functional_perf}. Importantly, the trends in energy errors and HOMO--LUMO gap errors evolve in opposite directions across the functional ladder, highlighting the distinct sensitivities of these quantities to the underlying exchange--correlation treatment.

\textbf{Energy Fidelity:}
Within QmDFT framework, LDA-RS yields the smallest deviations from the FCI-in-DFT reference, with an MAE of 0.092 mHa and an RMSE of 0.102 mHa as shown in Figure~\ref{fig:functional_perf}(a). Same behaviour is observed for the isomerization energies the results of LDA-RS are closer to theoretical coupled cluster singles doubles with perterbative triples (CCSD(t)) theory as well as experimental calculations from Roux et al.\cite{Roux2008} when compared to other functionals. As the fraction of global Hartree--Fock exchange increases, the deviations  become progressively larger, reaching maximum  for B3LYP MAE=1.259 mHa, RMSE=1.329 mHa. The results show that, for the embedding protocol considered here, lower-rung functionals exhibit higher variational optimization fidelity.

\textbf{Frontier-Orbital Gap estimate:} Contrastingly, the accuracy of frontier-orbital gaps relative to the experimental $0$--$0$ transition energies, quantified through $|E_{\mathrm{HL}}-E_{0-0}|$, follows the opposite trend as shown in Figure~\ref{fig:functional_perf}(b). The LDA-RS functional displays the largest deviations, with a gap MAE of $5.271\ \mathrm{eV}$ ($5.520\ \mathrm{eV}$ for the corresponding DFT calculations). Moving toward hybrid functionals substantially reduces these errors, with B3LYP producing the smallest QmDFT gap MAE of $0.471\ \mathrm{eV}$. This improvement is consistent with the reduced self-interaction error and the more balanced description of frontier-orbital energies provided by hybrid exchange-correlation treatments.

Together, the results show that superior performance for total energies does not translate into superior performance for frontier-orbital gaps. Instead, the observed trends demonstrate the property-dependent nature of functional performance within the embedding framework and the importance of assessing energetic and spectroscopic observables separately.

\section{Discussion}

\subsection*{Relevance to low-dimensional $\pi$-conjugated networks.}
Polycyclic aromatic hydrocarbons (PAHs), linear as well as fused are often used as molecular model systems for low-dimensional $\pi$-conjugated materials because they share characteristics with both small aromatic molecules and extended graphene nanoribbons~\cite{Hachmann2007acenes}. The gradual increase in conjugation across molecular size also makes it possible to examine how electronic properties evolve with system size.

Among these properties, the frontier-orbital gap is of particular interest due to its connection with optical and charge-transport phenomena~\cite{Coropceanu2007}. Reliable estimation of these gaps is therefore relevant to the computational study of organic semiconductors and related molecular electronic materials. At the same time, total energy accuracy remains important for assessing overall stability, so both aspects need to be considered together.

\subsection*{Energy accuracy versus frontier-orbital fidelity.}
The results present a trade-off that depends strongly on the choice of functional. The LDA-RS functional demonstrate close agreement with FCI-in-DFT benchmark in terms of total ground-state energies and thermochemical isomerization energies \cite{Roux2008} suggesting favorable error cancellation within the embedding procedure~\cite{Rossmannek2021}. However, the KS HOMO--LUMO gaps are consistently overestimated, which reflects the known limitations of local exchange functionals~\cite{Perdew1985}. In contrast, hybrid functionals provide significantly improved gap estimations. This behavior highlights the property-dependent nature of density-functional performance: functionals that reproduce total energies accurately do not necessarily provide the most reliable electronic gap estimates~\cite{Sun2016SCAN}.

We try to incorporate better exchange via Hybrid and range-separated hybrid functionals to provide a better description of frontier-orbitals and the improvement is visible in the KS HOMO--LUMO gap results getting closer to experimental 0-0 transition values in the uv-vis spectra. In particular, the B3LYP functional demonstrates close agreement with experimental values compared to other functionals. 

The CAM-B3LYP tuning procedure optimizes parameters to compress spectral property errors against a reference system, naturally shifts the total energy error profile. In this sense, calibration must be viewed as a deliberate, property-driven choice. The present results reinforce what has been observed in earlier studies: range-separated hybrids are often necessary for capturing the physics of extended $\pi$-systems and charge-transfer effects~\cite{Baer2010}.

\subsection*{Active space consistency and limitations.}
Using a fixed $(6e,6o)$ active space across all systems is mainly a practical choice, allowing for consistent comparison between different functionals. While this active space is sufficient to capture the main $\pi$-electron correlations, it is not fully sufficient to describe larger systems such as pentacene.

This limitation becomes visible in the residual discrepancies observed for the longer acenes in the LDA-RS functional error table and plot shown for pentacene Fig. \ref{fig:deltaE}. As the system size increases, correlation effects extend beyond the chosen active space to explain increase in the multireference character. That said, previous work ~\cite{Rossmannek2021}, which supports the approach taken here for smaller molecules.

\subsection*{Physical Justification of the Ground-State Proxy over Time-Dependent Formalisms}

The static Kohn-Sham (KS) orbital and the experimental 0-0 electronic transition energy represent two fundamentally different physical quantities. While, the actual experimental optical gap accounts for dynamic multi-configurational state-mixing, electron-hole screening, and geometry relaxation, the ground-state KS gap is a single-particle, static property. Our approach demonstrates that a calibrated, range-separated hybrid functional can be leveraged as a proxy in place of attempting to calculate this property via computationally heavy excited-state or time-dependent matrices like TD-DFT. By strategically optimizing the CAM-B3LYP parameters ($\mu=0.33$, $\alpha+\beta=0.28$), we effectively map the embedded frontier-orbitals to mirror experimental optical trends across single-transition dominated PAHs. This provides a cheap, computational alternative for estimation of experimental 0-0 electronic transition energy.

\subsection*{Implications for quantum computing workflows.}
From a computational perspective, the integration of VQE within the QmDFT framework is particularly interesting, as it provides a way to include correlated methods in quantum simulations~\cite{Peruzzo2014}. In this study, we focused on a noiseless statevector backend in order to separate methodological performance from hardware-related issues.

This simplification is important because current NISQ devices still face several limitations, including noise, limited qubit counts, and restricted circuit depth~\cite{Preskill2018}. These factors can significantly affect VQE performance and make convergence more difficult~\cite{Kandala2017}. By removing these effects, the present results give a clearer picture of what the method itself can achieve.

\subsection*{Quantum resource projections and scalability.}
One of the advantages of the embedding approach is the reduction in quantum resource requirements. For example, the $(6e,6o)$ active space of anthracene maps to roughly 10 qubits after Parity mapping and symmetry reductions~\cite{Bravyi2017}, which is within the qubit capacity of current quantum processors, although chemically accurate UCCSD circuit implementation challenging due to gate depth and noise. While a standard UCCSD ansatz is resource intensive due to the large number of single- and double-excitation operators, embedding-based smaller active space lowers practical circuit complexity as compared to full system. Embedding-based approaches therefore appear to be a promising way forward for studying larger $\pi$-conjugated systems on near-term quantum hardware.

This positions projection-based QmDFT as a bridge between DFT high-performance computing and near-term quantum hardware. Such reduced representations are particularly promising for low-dimensional materials, including graphene nanoribbons, where strong correlation and edge effects play a central role~\cite{Son2006}. Utilizing quantum embedding to address the multi-reference character of these compact active spaces represents a necessary step toward the modeling of larger, low-dimensional optoelectronic devices. Advances in ansatz design and error-mitigation strategies are likely to improve the practical performance of these methods in quantum simulations.

\subsection*{Overall perspective.}
Our results demonstrate that QmDFT delivers a highly customizable approach to electronic structure modeling. Instead of being locked into a rigid framework, one can tune the functional to emphasize the exact properties required for the application—be it precise total energies or accurate frontier-gaps. This property-driven flexibility is especially valuable for low-dimensional $\pi$-conjugated materials, where balancing both electronic aspects is critical, making it a far more robust alternative than sticking to a single default functional.

\section{Conclusion and Outlook}

This study examined a representative set of 10 PAHs including linear acenes and non-linear fused ring systems within a QmDFT embedding framework that supports advanced exchange--correlation functionals. Stable convergence is obtained via a combination of adaptive damping and DIIS acceleration, this allows evaluation of both total ground-state energies and KS HOMO--LUMO gaps within the same computational framework.

Comparing the QmDFT (VQE-in-DFT) energy results with FCI-in-DFT reference data demonstrates smallest errors for LDA-RS relative to the hybrid functionals. This is confirmed by thermochemical validation using isomerization energies of two isomeric pairs of PAHs anthracene/phenanthrene and tetracene/chrysene. HOMO--LUMO gap estimate of $E_{0-0}$ in contrast improve when hybrid and range-separated hybrid functionals are employed.

 CAM-B3LYP offers a balanced compromise that reproduces B3LYP's gap accuracy while reducing ground-state energy errors.
The results indicate that in QmDFT framework's preference to range separation and choice of functional depends on the target property. Local functionals perform well for energy calculations, whereas hybrids provide a more accurate description of frontier-orbital gaps. 

The active-space choice $(6e,6o)$ reduces the size of the quantum problem by restricting the treatment to  most relevant orbitals, which reduces the computational resource requirements for quantum simulations. This choice substantially reduces computational resources required VQE-based calculations bringing chemically relevant systems closer to the reach of near-term quantum hardware. This positions QmDFT as a practical link between traditional electronic structure methods and emerging quantum approaches.

Looking ahead, our framework can be extended to larger polycyclic systems, which have more prominent stronger multi-reference effects and edge-state physics. Addressing such challenging systems will require larger active space, better ansatz design, addition of error mitigation strategies for real Quantum device implementation. Progress in these directions will help improve estimation capability and broaden the applicability of these methods to low-dimensional materials and their use in molecular-scale optoelectronic and quantum technologies.

\section*{Acknowledgements}

The authors gratefully acknowledge the support of the AICTE Industry Fellowship Scheme for funding and facilitating this research. We are deeply grateful for the guidance and resources provided under this program, which directly enabled this research.

The authors also acknowledge the Centre for Development of Advanced Computing (C-DAC) and the National Supercomputing Mission (NSM) for providing computational resources and technical support. The contributions of interns Shreyas Kadam and Soham Phulare are sincerely appreciated.

N. Manglani would like to thank Dr. Sachin Nanawati for insightful discussions during the preparation of this manuscript. N. Manglani also gratefully acknowledges the hospitality at SRM University–AP and Prof. Mahesh Kumar Ravva for inspiring interest in polycyclic aromatic hydrocarbons (PAHs), which motivated this study.

\bibliography{paper2}    % NEW (no .bib extension)

\end{document}